\documentclass[iop,apj,numberedappendix,twocolappendix]{emulateapj}
\newcommand{\lea}{{\>\rlap{\raise2pt\hbox{$<$}}\lower3pt\hbox{$\sim$} \>}}
\newcommand{\gea}{{\>\rlap{\raise2pt\hbox{$>$}}\lower3pt\hbox{$\sim$} \>}}

\def\js{$j_\star$}
\def\Ms{$M_\star$}
\def\Us{$M_{\star}/L_K$}
\def\jsd{$j_{{\star}{\rm d}}$}
\def\jsb{$j_{{\star}{\rm b}}$}
\def\Msd{$M_{{\star}{\rm d}}$}
\def\Msb{$M_{{\star}{\rm b}}$}
\def\Usd{$(M_{\star}/L_K)_{\rm d}$}
\def\Usb{$(M_{\star}/L_K)_{\rm b}$}
\def\fd{$f_{\rm d}$}
\def\fb{$f_{\rm b}$}

\slugcomment{To appear in the Astrophysical Journal Letters}

\begin{document}

\title{ ANGULAR MOMENTUM AND GALAXY FORMATION REVISITED:\\
          EFFECTS OF VARIABLE MASS-TO-LIGHT RATIOS }

\author{S. Michael Fall\altaffilmark{1} and 
              Aaron J. Romanowsky\altaffilmark{2,3}
              }


\altaffiltext{1}{Space Telescope Science Institute,
         3700 San Martin Drive, Baltimore, MD 21218, USA}
\altaffiltext{2}{Department of Physics and Astronomy, 
         San Jos{\'e} State University,  One Washington 
         Square, San Jose, CA 95192, USA}
\altaffiltext{3}{University of California Observatories,
         1156 High Street, Santa Cruz, CA 95064, USA}

\begin{abstract}

We rederive the relation between the specific angular momentum \js\ and the mass \Ms\ of the stellar matter in galaxies of different morphological types.  
This is a revision of the \js--\Ms\ diagram presented in our recent comprehensive study of galactic angular momentum.  
In that work, we estimated \js\ from kinematic and photometric data that extended to large radii and \Ms\ from near-infrared luminosities $L_K$ with an assumed universal mass-to-light ratio \Us. 
However, recent stellar population models show large variations in \Us\ correlated with $B-V$ color. 
In the present work, we use this correlation to estimate \Us\ and hence \Ms\ from the measured $B-V$ and $L_K$.  
Our revised \js--\Ms\ diagram is similar to our previous one; both disk-dominated and elliptical galaxies follow nearly parallel sequences with $j_\star \propto M_{\star}^{\alpha}$ and $\alpha = 0.6 \pm 0.1$.  
However, the offset between the sequences is now a factor of about 5, some 30\% larger than before (and close to the offset found by Fall in 1983).  
Thus, our new results place even tighter constraints on the loss of specific angular momentum by galactic disks over their lifetimes.  

\end{abstract}


\section{INTRODUCTION}

Two of the most basic properties of galaxies are their angular momentum and mass.
These quantities are usually denoted by $J_{\star}$ and \Ms\ when they are restricted to the stellar matter in galaxies (i.e., excluding interstellar and dark matter).
The relationship between $J_{\star}$ and \Ms, or equivalently the one between $j_{\star} \equiv J_{\star} / M_{\star}$ and \Ms, provides both a physically motivated means of classifying galaxies and an important constraint on theories of galaxy formation \citep[hereafter F83]{1983IAUS..100..391F}.
The observed relation for disk-dominated galaxies has the form $j_{\star} \propto M_{\star}^{\alpha}$ with $\alpha \approx 0.7$, very close in both exponent and amplitude to the predicted relation if these galaxies retained most of the specific angular momentum they acquired by tidal torques in the early (linear and translinear) phases of their formation \citep{1980MNRAS.193..189F,1997ApJ...482..659D,1998MNRAS.295..319M}.
Elliptical galaxies follow a similar, nearly parallel, relation but offset to lower \js\ at each \Ms\ (by a factor of about 6 in the F83 study). 

We have recently completed a comprehensive reexamination of the galactic \js--\Ms\ diagram \citep[hereafter Paper I]{2012ApJS..203...17R}.  
The new study benefits especially from the large body of kinematic data for the outer parts of elliptical galaxies that has accumulated over the past three decades.  
In F83, the measurements for elliptical galaxies typically extended only to about their effective or half-light radii $R_{\rm e}$, and the rotation curves had to be extrapolated beyond this to estimate \js\ (assuming a constant rotation velocity for simplicity), whereas in Paper I, the measurements all extended to at least $\sim$$2R_{\rm e}$ and in a few cases beyond $6R_{\rm e}$, thus improving the accuracy with which \js\ could be estimated.  
This is important because, for elliptical galaxies, \js\ depends sensitively on the rotation curves beyond $R_{\rm e}$.  
The estimates of \js\ for galactic disks are more robust, in both F83 and Paper I, because they are less sensitive to the rotation curves at large radii and because the appropriate measurements (H$\alpha$ and 21 cm) have been available since the early 1980s.

The \js--\Ms\ diagram in Paper I is similar to the one in F83 except that the offset between the disk-dominated and elliptical galaxies has shrunk to a factor of 3--4.  
This difference is not primarily due to improvements in the kinematic data and hence in the estimates of \js\ for elliptical galaxies, which turn out to be remarkably similar, on average, in both studies.  
Rather, the different offsets in the \js--\Ms\ relations are due largely to the different ways \Ms\ was estimated.  
In F83, \Ms\ was computed from $B$-band luminosities and mass-to-light ratios, using the correlation between $M_{\star}/L_B$ and $B-V$ from stellar population models \citep{1981MNRAS.194...63T}; in Paper I, \Ms\ was computed from $K$-band luminosities and mass-to-light ratios, assuming a universal $M_{\star}/L_K = 1$ (in solar units) for simplicity.  
Until recently, it was widely believed that the variations in \Us\ were weak enough to be neglected in practice \citep{2003ApJS..149..289B}.
We have taken this luminosity-based approach because we are primarily interested in the \js--\Ms\ relations for the stellar matter in galaxies, and because we want to avoid the spurious correlations that can arise when estimating both \js\ and \Ms\ from the same kinematic data.   

In this Letter, we return to the \js--\Ms\ relation for galaxies of different morphological types.  
Our focus here is on how much the $K$-band mass-to-light ratio varies among galaxies and how this affects the offset between disk-dominated and elliptical galaxies in the \js--\Ms\ diagram.   
We first compile results from several stellar population models to show how \Us\ correlates with $B-V$ color (Section 2).  
We then use the measured $B-V$ colors of galaxies, corrected for extinction, to estimate \Us\ and hence \Ms\ from the measured $L_K$.  
This leads to our revised \js--\Ms\ diagram (Section 3).  
Finally, we discuss some of the implications of our results for understanding the origin and evolution of galaxies (Section 4).  
The work presented here is based heavily on Paper I, which we urge readers to consult for both general and specific information on this topic.

\section{MODEL \Us\ VARIATIONS}

The mass-to-light ratios of stellar populations are correlated with their colors because both of these quantities depend on the star-formation histories of the populations.
A galaxy with a high proportion of young stars will have a low mass-to-light ratio and blue colors, and vice versa.
This has been known for a long time for optical mass-to-light ratios and optical colors: both $M_{\star}/L_B$ and $M_{\star}/L_V$ rise steeply with $B-V$ (e.g., \citealt{1981MNRAS.194...63T}).
Recent stellar population models indicate that near-infrared mass-to-light ratios also have strong correlations with optical colors \citep{2001ApJ...550..212B,2009MNRAS.400.1181Z,2013MNRAS.430.2715I}, a result we confirm here.
We focus on the correlation between \Us\ and $B-V$ because we have measurements of $L_K$ and $B-V$ for most of the galaxies in our sample. 

\begin{figure}
\centering{
\includegraphics[width=3.3in]{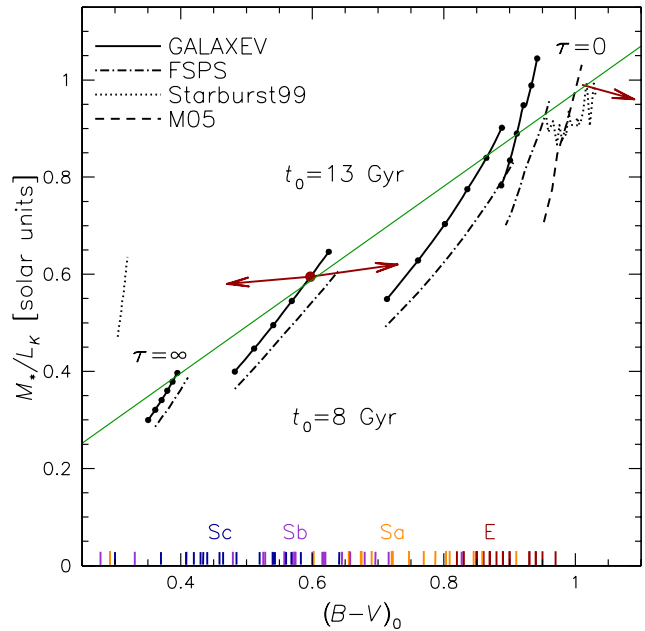}
}
\caption{$K$-band mass-to-light ratio plotted against extinction-free $B-V$ color for stellar population models with exponential SFRs, several decay times $\tau$, and initial ages that vary continuously from $t_0 = 8$~Gyr to $t_0 = 13$~Gyr.  
The models are: GALAXEV and FSPS with $\tau=0$, 2~Gyr, 5~Gyr, $\infty$; Starburst99 with $\tau=0$, $\infty$; and M05 with $\tau=0$.
The arrows show the effects of $\pm 0.6$ mag of extinction in the $B$~band (two-headed arrow) and an increase in the metallicity from $Z_\odot$ to $2.2Z_\odot$ (one-headed arrow).
The diagonal green line represents our adopted correlation between \Us\ and $(B-V)_0$ (Equation~(1)).
The tick marks at the bottom of the figure indicate the extinction-corrected $B-V$ colors of the galaxies in our sample.}\label{fig:fig1}
\end{figure}

In Figure~\ref{fig:fig1}, we plot \Us\ against $B-V$ for several stellar population models with different star-formation histories.\footnote
{Throughout this Letter, \Ms\ stands for current (not initial) stellar mass, including remnants, and $B-V$ is on the Vega magnitude system.}
The models are: GALAXEV (the most recent version of the \citet{2003MNRAS.344.1000B} model, called CB$^\star$ and described in \citet{2012MNRAS.426.2300L}), FSPS \citep[version 2.3]{2009ApJ...699..486C,2010ApJ...712..833C}, Starburst99 \citep{2005ApJ...621..695V}, and  M05 \citep{2005MNRAS.362..799M}.
The curves are based on models with a normal initial mass function (IMF; 
\citealt{2003PASP..115..763C}), solar metallicity ($Z = 0.017$), and no extinction (hence the subscript 0 on $B-V$).
In all cases, the star-formation rate (SFR) declines exponentially: ${\rm SFR} \propto \exp(-t/\tau)$.
The adopted decay times range from $\tau=0$, an instantaneous burst, most relevant to galactic bulges and elliptical galaxies, to $\tau = \infty$, a constant SFR, most relevant to galactic disks.
For the GALAXEV and FSPS models, the decay times are $\tau=0$, 2~Gyr, 5~Gyr, and $\infty$; for the Starburst99 models, $\tau=0$ and $\infty$; and for the M05 model, $\tau=0$.
Each curve represents a continuous sequence of initial ages, from $t_0 = 8$~Gyr ($z \approx 1$) to $t_0 = 13$~Gyr ($z \approx 7$--12), with a fixed value of $\tau$.

Figure~\ref{fig:fig1} shows that \Us\ generally increases with $(B-V)_0$.
The fact that the curves are mostly parallel to each other indicates that the primary link between \Us\ and $(B-V)_0$ is their mutual dependence on the ratio of the present SFR to the past-average SFR, $b =  (t_0/\tau)/[\exp(t_0/\tau) - 1]$ for exponential SFRs.
This correlation is approximate, not exact; models with different star-formation histories have a range of \Us\ at the same $(B-V)_0$.
As Figure~\ref{fig:fig1} shows, there is generally good agreement among different models with the same values of $\tau$ and $t_0$.
The discrepancies are likely attributable to the different treatments of thermally pulsing stars on the asymptotic branch (TP-AGB stars), and are mostly smaller than the uncertainties caused by the unknown star-formation histories of galaxies. 
The arrows in Figure~\ref{fig:fig1} show the results of $\pm 0.6$ mag of extinction in the $B$~band (two-headed arrow) and an increase in the metallicity from $Z_\odot$ to $2.2Z_\odot$ (one-headed arrow).
Extinction, as expected, has a negligible effect on \Us.
Metallicity matters more, but not as much as the star-formation history.  
 
We adopt the following correlation between $K$-band mass-to-light ratio and extinction-free $B-V$ color: 
\begin{equation}
M_{\star}/L_K = 0.96 (B-V)_0  + 0.01 \quad ({\rm solar \,\, units}).
\end{equation}
This formula, plotted as the diagonal straight line in Figure~\ref{fig:fig1}, is meant to approximate the overall trend from contemporary stellar population models with diverse star-formation histories.
The slope in Equation~(1) is uncertain by $\pm0.2$.
Our adopted correlation is broadly consistent with results from other recent studies based on stellar population models \citep{2009MNRAS.400.1181Z,2013MNRAS.430.2715I}.
Equation~(1), while perhaps not definitive, is a significant improvement on the universal \Us\ adopted in Paper~I (motivated largely by Table~7 of \citealt{2003ApJS..149..289B}).
Two alternative formulae that approximate the curves in Figure~\ref{fig:fig1} at least as well as Equation~(1) are $M_{\star}/L_K = (B-V)_0$ and $\log(M_{\star}/L_K) = 0.75 [(B-V)_0 - 1.0]$.
We have checked that the main results of this Letter (Figures~\ref{fig:fig2} and~\ref{fig:fig3}) are not sensitive to the exact form of the correlation between $M_{\star}/L_K$ and $(B-V)_0$.

In principle, we could use dynamical measurements to test Equation~(1).
The initial results of the DiskMass project suggest a fairly strong correlation between \Us\ and optical colors in galactic disks, but possibly with a lower normalization than adopted here \citep{2011ApJ...739L..47B,2011ApJ...742...18W}.
On the other hand, \citet{2009MNRAS.400.1665W} find a weaker correlation and a higher normalization  (see their Figure~9).
Another concern is that the IMF may vary among and within galaxies.
There is indirect evidence that the central parts of the reddest, most metal-rich, elliptical galaxies have steeper-than-normal IMFs, with \Us\ roughly twice that predicted by Equation~(1) for $(B-V)_0 \approx 1$ (see
\citealt{2013ApJ...765....8T} for a summary).
A dependence of the IMF slope on metallicity could also cause radial gradients in \Us\ within galaxies, although it is not yet clear how strong such variations might be \citep{2012ApJ...760...71C,2013ApJ...763..110K}.

\section{REVISED \js--\Ms\ DIAGRAM}

We now use the results from the previous section to revise our \js--\Ms\ diagram from Paper~I as follows.
We subdivide galaxies into their disk (d) and bulge (b) components, with specific angular momenta \jsd\ and \jsb, mass-to-light ratios \Usd\ and \Usb, and fractions \fd\ and \fb\ of the total $K$-band luminosity $L_K$.
Then the total angular momentum and total mass of a galaxy are given by
\begin{equation}
J_{\star} = j_{{\star}{\rm d}} M_{{\star}{\rm d}} +  j_{{\star}{\rm b}} M_{{\star}{\rm b}},
\end{equation}
\begin{equation}
M_{\star} = M_{{\star}{\rm d}} + M_{{\star}{\rm b}},
\end{equation}
\begin{equation}
M_{{\star}{\rm d}} = f_{\rm d} (M_{\star}/L_K)_{\rm d} L_K,
\end{equation}
\begin{equation}
M_{{\star}{\rm b}} = f_{\rm b} (M_{\star}/L_K)_{\rm b} L_K.
\end{equation}

In Paper~I, we estimated these quantities for a sample of 95 galaxies (57 spirals, 15 lenticulars, and 23 ellipticals) covering the full range of disk-to-bulge ratios and morphological types.
We derived the specific angular momenta \jsd\ and \jsb\ from the observed rotation curves and surface brightness profiles, after deprojection by individual inclination angles for spiral galaxies (with known orientations) and by the median inclination angle for lenticular and elliptical galaxies (with random orientations).
It is important to note that \jsd\ and \jsb\ are independent of \Usd\ and \Usb\ provided only that radial gradients in these mass-to-light ratios are negligible within each component. 
The luminosity fractions \fd\ and \fb\ were taken from the two-dimensional decomposition of $r$-band images into disk and bulge components by \citet{1986AJ.....91.1301K,1987AJ.....93..816K,1988AJ.....96..514K} and others; we neglect the small differences between these fractions in the $r$ and $K$ bands. 
The total $K$-band luminosities were derived from $K_{20}$ isophotal magnitudes in the Two Micron All Sky Survey and our extrapolations based on the surface brightness profiles from Kent and others.
We then computed $j_{\star} \equiv J_{\star} / M_{\star}$ and $M_{\star}$ from Equations~(2)--(5) with $(M_{\star}/L_K)_{\rm d} = (M_{\star}/L_K)_{\rm b} = 1$.

Here, we repeat these calculations but with different mass-to-light ratios for the disk and bulge components based on our adopted correlation between \Us\ and $(B-V)_0$. 
HyperLeda \citep{2003A&A...412...45P} lists $(B-V)_0$ for all but 13 galaxies in our sample (mostly spirals), indicated by the tick marks at the bottom of Figure~\ref{fig:fig1}; these are corrected for both foreground and internal extinction.
Since we do not have colors measured separately for the disk and bulge components, we proceed as follows.
For bulges, we adopt the median color $(B-V)_{0{\rm b}} = 0.87$ for the elliptical galaxies in our sample and the corresponding $(M_{\star}/L_K)_{\rm b} = 0.84$ from Equation~(1).
This is reasonable because the colors of old stellar populations are affected more by their metallicities than their star-formation histories.
For the disks of spiral galaxies, the converse is true; their colors depend mainly on their present SFRs, rather than their metallicities.
Thus, we estimate the color of each disk $(B-V)_{0{\rm d}}$ from the color of the whole galaxy $(B-V)_0$, the adopted color of the bulge $(B-V)_{0{\rm b}}$, and the luminosity fractions \fd\ and \fb.  
We then derive \Usd\ from $(B-V)_{0{\rm d}}$ and Equation~(1).
For lenticular galaxies, unfortunately, we do not have reliable estimates of the disk and bulge properties separately.
Thus, we simply adopt the overall \js\ from Paper~I and derive the overall \Ms\ from $L_K$ and \Us, the latter computed from the overall $(B-V)_0$ and Equation~(1).
We have tested this procedure on spiral galaxies with small disk-to-bulge ratios and find that it is accurate at the 5\% level.

\begin{figure}
\centering{
\includegraphics[width=3.3in]{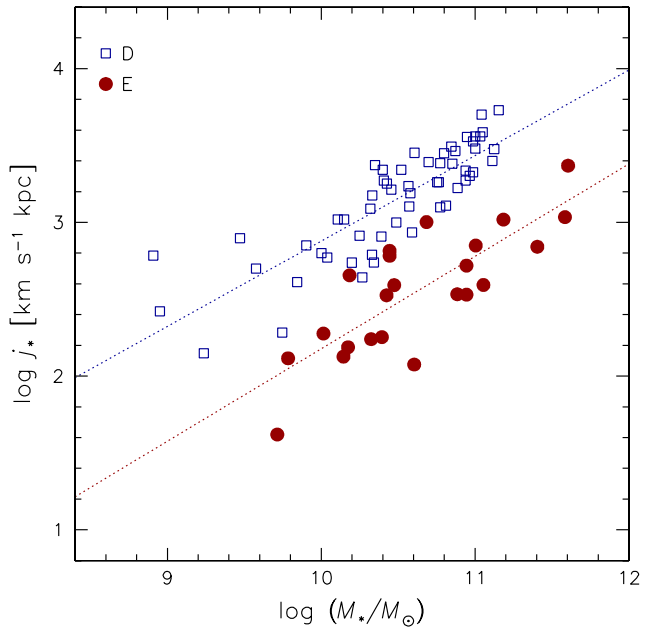}
}
\caption{Specific angular momentum plotted against mass for the disks alone of spiral galaxies (open squares) and for elliptical galaxies (filled circles). 
The dotted lines represent $j_{\star} \propto M_{\star}^{\alpha}$ with $\alpha = 0.6$.}\label{fig:fig2}
\end{figure}

In Figure~\ref{fig:fig2}, we plot specific angular momentum against mass for the disks {\it alone} of spiral galaxies (i.e., \jsd\ versus \Msd) and for elliptical galaxies (i.e., \jsb\ versus \Msb).
This is our most robust result because it avoids the complications and uncertainties in apportioning angular momentum and mass between the disks and bulges of composite galaxies.
Indeed, the vertical locations ($\log j_{\star}$) of the points in Figure~\ref{fig:fig2} here are the same as in the right panel of Figure~14 in Paper~I; only the horizontal locations ($\log M_{\star}$) have changed as a result of our revised mass-to-light ratios. 
The disks have moved leftward significantly, while the ellipticals have moved only slightly.
As before, the disks and ellipticals follow nearly parallel sequences: $j_{\star} \propto M_{\star}^{\alpha}$ with $\alpha = 0.6 \pm 0.1$ in both cases.
However, the gap between these sequences has increased from a factor of $3.8 \pm 0.5$ in Paper~I to $4.8 \pm 0.6$ here.
If elliptical galaxies were to have steeper IMFs than the disks of spiral galaxies, the gap between the two sequences in Figure~\ref{fig:fig2} would increase, and if the IMF were to depend on metallicity, the exponent $\alpha$ of the \js--\Ms\ relation might also be affected.

\begin{figure}
\centering{
\includegraphics[width=3.3in]{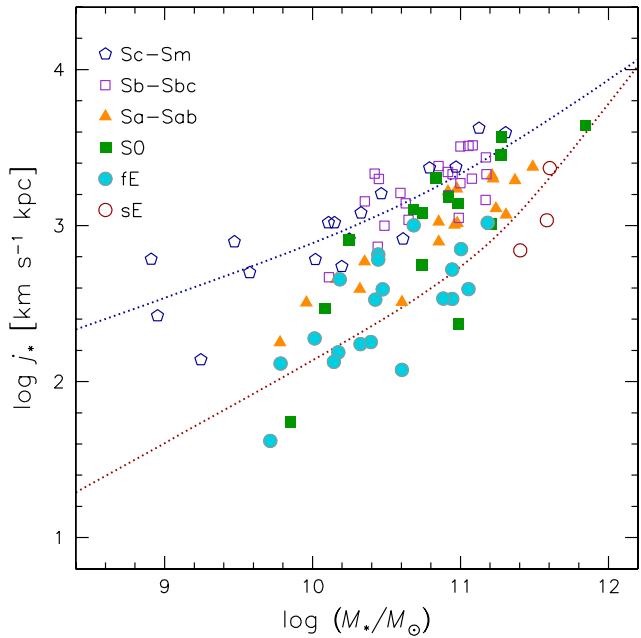}
}
\caption{Specific angular momentum plotted against mass for entire galaxies (disk plus bulge) of different morphological types. 
The dotted curves represent our simple model of variable retention of specific angular momentum in a $\Lambda$CDM cosmogony with retained fractions $f_j = 0.8$ (upper curve) and $f_j = 0.1$ (lower curve).}\label{fig:fig3}
\end{figure}

Figure~\ref{fig:fig3} shows our results for entire galaxies (disk plus bulge) of all morphological types with \js\ and \Ms\ computed by the full procedure described above.
This updates the left panel of Figure~14 in paper~I for variable $M/L_K$. 
The revised \js--\Ms\ diagram confirms our earlier conclusion that galaxies of intermediate disk-to-bulge ratios and morphological types fill in the gap between the sequences of pure disks (late-type spiral galaxies) and pure bulges (elliptical galaxies).
The full spread in \js\ at each \Ms\ is about 30\% larger here than in Paper~I, again because the disks have moved farther to the left than the bulges have.  
The dashed curves in Figure~\ref{fig:fig3} were derived from our simple model of variable retention of specific angular momentum during galactic formation and evolution in the standard cold dark matter ($\Lambda$CDM) cosmogony, with retained fractions $f_j = 0.8$ for disks and $f_j = 0.1$ for ellipticals (see Section~6 of Paper~I).
Evidently, this model matches the data remarkably well.
For disks, our new estimate of $f_j$ is about 30\% larger than in Paper~I, while for ellipticals, it is essentially the same.

\section{CONCLUSIONS}

The results presented here reinforce our previous conclusions that galaxies of all morphological types lie along nearly parallel sequences with exponents $\alpha \approx 0.6$ in the \js--\Ms\ diagram (F83; Paper~I).  
Our revised \js--\Ms\ relation is based on the best available kinematic and photometric data to estimate \js\ and $L_K$ (taken from Paper~I) and contemporary stellar population models to estimate \Us\ from $(B-V)_0$ (Figure~\ref{fig:fig1}).  
We find that the offset between galactic disks and elliptical galaxies is now a factor of about 5 (Figure~\ref{fig:fig2}), slightly lower than the factor of about 6 found by F83 and slightly higher than the factor of 3--4 found in Paper~I.  
These differences are small enough that our basic interpretation of the \js--\Ms\ relation remains 
qualitatively intact, with only minor quantitative revisions.
Thus, we give only a brief summary here and refer the interested reader to Paper~I for a more 
complete discussion. 

In Paper I, we showed that the positions of galaxies in the \js--\Ms\ diagram are correlated with their positions in the two-dimensional spaces of disk-to-bulge ratio versus mass and morphological type versus mass (confirming a conjecture by F83).  
Thus, the description of galaxies in terms of \js\ and \Ms\ may be regarded as a physically motivated alternative to the more familiar classification schemes based on disk-to-bulge ratio or morphological type.  
Our new results reinforce this conclusion; we now find slightly stronger correlations than in Paper~I between disk-to-bulge ratio, morphological type, and specific angular momentum.

An advantage of the \js--\Ms\ approach is that the ``initial'' conditions in this diagram are known with high confidence from cosmological $N$-body simulations.  
In $\Lambda$CDM and other hierarchical models of structure formation, protogalaxies acquire their angular momenta $J$ by tidal torques, which leads to an approximate lognormal distribution of the spin parameter $\lambda$, independent of mass $M$ (e.g., \citealt{1987ApJ...319..575B,1992ApJ...399..405W,1996MNRAS.281..716C}).
This in turn corresponds to a lognormal distribution of $j \equiv J/M$ at each $M$ with a dispersion $\sigma(\ln{j}) \approx 0.5$ and a mean relation $j \propto M^{\alpha}$ with $\alpha \approx 2/3$. 
The baryonic matter in galaxies should initially follow a similar distribution, since it must have experienced the same tidal torques as the dark matter before virialization, radiative dissipation, and other non-linear effects \citep{1980MNRAS.193..189F,1997ApJ...482..659D,1998MNRAS.295..319M}.
It is therefore significant that the observed \js--\Ms\ relations for galaxies have nearly the same exponent $\alpha$ as the predicted relation for dark halos (as noted by F83).  

The locations of galaxies in the \js--\Ms\ diagram can be altered over their lifetimes by a variety 
of processes.  
These include the exchange of angular momentum between baryons and dark matter, merging, stripping, inflow, outflow, and star formation, each of which may be represented by a vector ($\Delta j_{\star}$, $\Delta M_{\star}$).  
The sum of these vectors over all processes must then map the predicted halo sequence in the 
$j$--$M$ diagram into one of the observed galactic sequences.  
The fact that these sequences are nearly parallel to each other means that the net fractional changes in \js\ must be almost independent of \Ms\ even though the fractional changes in \js\ and \Ms\ may be large.  
This is a powerful constraint on all models of galactic formation and evolution. 
In the idealized case that \js\ is reduced while \Ms\ remains constant, we find that galactic disks and elliptical galaxies must have retained fractions $f_j \approx 0.8$ and 0.1, respectively, of their initial specific angular momenta (Figure~\ref{fig:fig3}).

Many of the processes affecting galaxies will change $j$ and $M$ simultaneously.  
For example, merging tends to decrease $j$ on average (by the vector cancellation of spins) while increasing $M$.  
If elliptical galaxies formed mainly by mergers, this could account for much of their offset from disk-dominated galaxies in the \js--\Ms\ diagram.  
Outflows, which decrease $M$, can increase or decrease $j$ depending on whether they 
remove material predominantly from the inner or outer parts of galaxies.  
This would broaden the \js--\Ms\ relation, possibly by the observed amount. 
Another way of explaining the observed \js--\Ms\ relation is in terms of the partial collapse 
of baryons within their dark halos (called ``collapse bias'' in Paper I).  
It is not yet clear how all these processes cooperate to produce the galaxies we observe.  
But as we have emphasized here and in more detail in Paper I, the options are severely constrained by the predicted initial conditions and observed galactic sequences in the \js--\Ms\ diagram.

\acknowledgements

We thank Eric Bell, Gustavo Bruzual, St\'ephane Charlot,
Charlie Conroy, Claus Leitherer, and Stefano Zibetti for helpful 
discussions and for providing their models in electronic form. 
This work was supported in part by NSF grant AST-0909237.

\end{document}